# Plasma vortices driven by magnetic torsion generated by electric currents in non-magnetic planetary wakes

Hector Javier Durand-Manterola, Alberto Flandes, Hector Perez-de-Tejada

Departamento de Ciencias Espaciales, Instituto de Geofısica, Universidad Nacional Autonoma de Mexico, Coyoacan 04510, Mexico City, Mexico.
E-mails:
durand-manterola@igeofisica.unam.mx
alberto_flandes@igeofisica.unam.mx

Abstract
In non-collisional magnetized astrophysical plasmas, vortices can form as it is the case of the Venus plasma wake where Lundin et al. (2013) identified a large vortex through the integration of data of many orbits from the Venus Express (VEX) spacecraft. On the one hand, our purpose is to develop a theoretical foundation in order to explain the occurrence and formation of vortices in non- collisional astrophysical plasmas. On the other hand, to apply the latter in order to study the vorticity in the wakes of Venus and Mars. We introduce two theorems and two corollaries, which may be applicable to any non-collisional plasma system, that relate the vorticity to electromagnetic variables such as the magnetic field and the electric current density. We also introduce a toy vortex model for the wakes of non-magnetized planetary bodies. From the proposed theorems and model and using magnetic data of the VEX and the Mars Global Surveyor (MGS) spacecraft, we identify vortices in the wakes of Venus and Mars in single spacecraft wake crossings. We also identify a spatial coincidence between current density and vorticity maxima confirming the consistency of our theorems and model. We conclude that vortices in non-collisional magnetized plasmas are always linked to electric currents and that both vortices and currents always coexist. This suggests that the mechanism that produces this type of vortices is the mutual interaction between the electric current and the magnetic field, that to a first approximation is explained considering that plasma currents due to a non-zero net charge density induce magnetic fields that modify the existing field and also produce a helical field configuration that drives charged particles along helical trajectories.

Highlights
• We introduce two theorems and two corollaries, applicable to any non-collisional plasma system, that relate the vorticity to electromagnetic variables.
• We also introduce a toy vortex model for the wakes of non-magnetized planetary bodies.

- We identify a spatial coincidence between current density and vorticity maxima confirming the consistency of our theorems and model.

1. Introduction

Since the 1950s, it has been known that vortices can form in plasmas as in neutral gases (Buneman, 1958). However, vortices in non-collisional plasmas in astrophysical environments are of more recent discovery. Early theoretical works like Banks et al. (1981) recognized the importance of vorticity in space plasmas as those found in the ionosphere of Venus and near comets. Perez-de-Tejada et al. (1982) pointed out that the Pioneer Venus spacecraft data suggested the presence of a vortex-like structure in the Venus plasma wake, which could not be confirmed at the time, because of the scarce data available. Almost 30 years later, Lundin et al. (2013) reported the detection of a vortex in the Venus plasma wake through the integration of data of many orbits from the Venus Express (VEX) spacecraft. This latter structure was observed in the solar wind proton component and in the dragged planetary oxygen ions. Measurements showed that the vortex rotated in the clockwise direction -looking from Venus towards the wake- and that its axis was parallel to the Sun-Venus direction, while it extended along the wake. Another type of astrophysical vortices is given by Ruhunusiri et al. (2016), who report the observation of ion vortical structures in the Martian sheath-ionosphere boundary.

In their work, Lundin et al. (2013) posed some questions that still remain unanswered and that we consider relevant and list here: What role does the solar wind and induced magnetic fields play in the process? What determines the vortex rotation direction? Is the Venus large-scale vortex a steady-state phenomenon? If not, what determines its variability?

An important part of the goal of the current work is devoted to answer the above questions. In order to do so, our approach concentrates on the analysis of vorticity, which is a variable that determines the rotation to which a fluid is subject. In this context, any circular or rotating flow that has non-zero vorticity can be considered a vortex. Another important issue is whether these vortices in astrophysical plasmas are always purely hydrodynamic phenomena. As a simple test, let us take the solar wind flow interacting with a non- magnetized planet and assume a velocity field with axial symmetry -in cylindrical coordinates- as:

$$\mathbf{u} = \big(u_r(r,z), 0, u_z(r,z)\big) \qquad (1)$$

where r is the radial coordinate measured from the axis of the wake; z the coordinate parallel to the axis along the Sun-body direction; $u_r$ is the radial solar wind speed and $u_z$ is the velocity component parallel to the wake axis. Directly, the vorticity, $\boldsymbol{\omega} = \nabla \times \mathbf{u}$, that this type of flow would have would be:

$$\omega = \frac{1}{r}\begin{vmatrix} \hat{r} & r\hat{\phi} & \hat{k} \\ \frac{\partial}{\partial r} & \frac{\partial}{\partial \phi} & \frac{\partial}{\partial z} \\ u_r(r,z) & 0 & u_z(r,z) \end{vmatrix} = -\left(\frac{\partial u_z}{\partial r} - \frac{\partial u_r}{\partial z}\right) r\hat{\phi} \qquad (2)$$

According to the latter, the axis of the vortex would be perpendicular to the flow, which could give us a hint to help understand the return flows reported by Masunaga et al. (2019); Bader et al. (2019); Lundin et al. (2011) and Perez-de-Tejada et al. (2019). However, in the case of the vortex reported by Lundin et al. (2013) the axis of the vortex, and therefore its vorticity, is parallel to the flow. This suggests that the vortex in the wake of Venus is not only driven by hydrodynamic mechanisms, but that other mechanisms seem to be involved and must be sought. In order to help answer the above questions and based on the latter, first, in Section 2, we propose two theorems and two corollaries based on first principles, which justify those electric currents produce vorticity in non-collisional magnetized plasmas which can be expressed as function of the same electric current or the magnetic field. We also propose a toy model that explains in a simple way, vortices in wakes of non-magnetized planetary bodies. In Section 3, we show the results of applying the model developed in Section 2 to data of the VEX and the MGS spacecraft. In Section 4 we discuss our results.

## 2. Vorticity in magnetized plasmas
### 2.1. Theorems.

Let S, our system of study, be a portion of a non-collisional magnetized moving plasma, with only two species of charge carriers: protons and electrons, Let $n_p$ and $n_e$, the proton and electron plasma density, such that $n_p = n_e = n$. Also, let $u_p$ and $u_e$, the proton and electron velocity, such that $u_p \neq u_e$ and $u_p - u_e = u$.

### 2.1.1. Theorem I
Hypothesis
Let $\omega$ be the plasma vorticity, and J, the electric current density
Thesis:
The vorticity in S is expressed by:

$$\omega = \frac{\nabla \times J}{nq} - \frac{\nabla n \times u}{n} \quad (3)$$

Proof:
By definition (Reitz et al. 2008, p. 166) the current density is:

$$\mathbf{J} = \sum_k n_k q_k \mathbf{u}_k \quad (4)$$

where the index k indicates the different charged species, $n_k$ is the numerical density, $q_k$ and $u_k$ are the charge, and the velocity of the k-th species of charge carriers.
In terms of the proton and electron density as well as their speeds, Equation 4 can be written as:

$$\mathbf{J} = q(n_p \mathbf{u}_p - n_e \mathbf{u}_e) \quad (5)$$

or in terms of the plasma density, n, and given that $u_p - u_e = u$:

$$\mathbf{J} = qn\mathbf{u} \quad (6)$$

Applying the curl to Equation 6, we have:

$$\nabla \times \mathbf{J} = nq \, \nabla \times \mathbf{u} + q \nabla n \times \mathbf{u} \quad (7)$$

From the definition of the vorticity ($\omega = \nabla \times u$), then:

$$\nabla \times \mathbf{J} = nq\boldsymbol{\omega} + q \nabla n \times \mathbf{u} \quad (8)$$

Therefore, solving for ω:

$$\boldsymbol{\omega} = \frac{\nabla \times \mathbf{J}}{nq} - \frac{\nabla n \times \mathbf{u}}{n} \quad (3)$$

QED

Equation 3 states that, in non-collisional magnetized plasmas, a purely hydrodynamic variable, such as the vorticity, can be expressed, not only in terms of plasma properties as the velocity, and the plasma density, but in terms of electrical variables such as the current density.

2.1.2. Corollary I
Hypothesis
The plasma density, n, is constant in space or its variations are small enough.
Thesis
From Theorem 1, it follows that:

$$\boldsymbol{\omega} = \frac{\nabla \times \mathbf{J}}{nq} \quad (9)$$

Proof:
If in the system n is either constant or its variations are small enough in space, we can assume either that $\nabla n = 0$ or $\approx 0$, and then in Equation 3 the second term vanish, and equation (3) converts in equation (9). QED.

In the end, Equation 9 states that the most relevant term in Equation 3, in terms of vorticity, is the one which carries the current density.

2.1.3. Theorem II
If Theorem 1 holds, the vorticity can also be expressed as function of the magnetic field as:

$$\omega = \frac{\Box^2 B}{\mu_0 n q} \quad (10)$$

where $\Box^2$ represents the D'Alambertian or d'Alambert operator.

Proof:
Let depart from Ampere-Maxwell's Law:

$$\nabla \times B = \mu_0 J + \varepsilon_0 \mu_0 \frac{\partial E}{\partial t} \quad (11)$$

Solving J:

$$J = \frac{1}{\mu_0} \nabla \times B - \varepsilon_0 \frac{\partial E}{\partial t} \quad (12)$$

and applying the curl to Equation 12:

$$\nabla \times J = \frac{1}{\mu_0} \nabla \times (\nabla \times B) - \varepsilon_0 \nabla \times \frac{\partial E}{\partial t} \quad (13)$$

Recalling that $\nabla \times (\nabla \times B) = \nabla(\nabla \cdot B) - \nabla^2 B$, where $\nabla \cdot B = 0$, and that spatial coordinates and time are independent variables and, therefore the curl and the time derivative can be exchanged:

$$\nabla \times J = -\frac{1}{\mu_0 n q} \nabla^2 B - \frac{\varepsilon_0}{n q} \frac{\partial}{\partial t} \nabla \times E \quad (14)$$

From Faraday's induction law:

$$\nabla \times J = -\frac{1}{\mu_0 n q} \nabla^2 B + \frac{\varepsilon_0}{n q} \frac{\partial^2 B}{\partial t^2} \quad (15)$$

By corollary I:

$$\omega = -\frac{1}{\mu_0 n q} \nabla^2 B + \frac{\varepsilon_0}{n q} \frac{\partial^2 B}{\partial t^2} \quad (16)$$

In terms of the speed of light, $c^2 = 1/\mu_0 \varepsilon_0$, is:

$$\omega = \frac{1}{\mu_0 n q} \left( \frac{1}{c^2} \frac{\partial^2 B}{\partial t^2} - \nabla^2 B \right) \quad (17)$$

where the term inside the parenthesis is the d'Alembertian of B. Therefore

$$\omega = \frac{\Box^2 B}{\mu_0 n q} \quad (10)$$

QED

### 2.1.4. Corollary II
If the system is in a steady state, the vorticity can be expressed as:

$$\omega = -\frac{\nabla^2 B}{\mu_0 n q} \quad (18)$$

Proof:
Equation 18 follows directly from Equation 17 assuming that the system is in steady state, that is, $\partial B/\partial t = 0$ and thus $\partial^2 B/\partial t^2 = 0$ as well. After this assumption, only the term $-\nabla^2 B$ remains.

QED

### 2.2. Toy vortex model for the plasma wakes of non-magnetic planetary bodies.
Let us suppose a cylindrical axisymmetric wake with an axisymmetric current as well. Based, either on the Biot-Savart's law or the Ampere's Law (e.g., Reitz et al., 2008, pp 173), the magnetic field generated by the proposed current in cylindrical coordinates, may be written as:

$$\overline{B_c} = \frac{\mu_0 i}{2\pi r} \hat{\varphi}_0 \quad (19)$$

where i is the current intensity within a radial distance r at which the field is evaluated. In terms of the current density, j, and assuming a circular cross-section area:

$$i = \int_0^R 2\pi r j(r) dr \quad (20)$$

Where R is the radius of the wake.
Let the current density be described by a function of the type:

$$j = j_0 \exp(-Dr^2) \quad (21)$$

This function guarantees that the current density has its maximum value, $j_0$, at the axis of the wake and declines with the distance to the axis. D is a constant that determines the width of the current density peak, whose value is adjusted so that j ≈ 0 at r = R.
Therefore

$$i = \int_0^R 2\pi j_0 r \exp(-Dr^2) dr \quad (22)$$

If $u = -Dr^2$ and $du = -2Dr dr$, Equation 22 turns into:

$$i = -\frac{\pi j_0}{D} \int_0^u exp(u) du \quad (23)$$

And integrating:

$$i = \frac{\pi j_0}{D}[1 - exp(-DR^2)] \quad (24)$$

Substituting Equation 24 into Equation 19:

$$\bar{B}_c = \frac{\mu_0}{2R}\frac{j_0}{D}[1 - exp(-DR^2)]\widehat{\varphi_0} \quad (25)$$

Now, let the draped magnetic field flux around the planet be:

$$\bar{B}_d = B_{rd}\hat{r}_0 + B_{\varphi d}\widehat{\varphi_0} + B_{xd}\hat{\imath} \quad (26)$$

Contrary to the convention, we use x instead of z as the third cylindrical component, in order to be consistent with the spacecrafts data that we use in Section 3, where the negative x-direction points along the wake.
With the above considerations, the total magnetic field is:

$$\bar{B} = \bar{B}_d + \bar{B}_c = B_{rd}\hat{r}_0 + \left(B_{\varphi d} + \frac{\mu_0}{2R}\frac{j_0}{D}[1 - exp(-DR^2)]\right)\widehat{\varphi_0} + B_{xd}\hat{\imath} \quad (27)$$

Nevertheless, if we assume that the dominant draped field is oriented along the x-axis, which is an appropriate assumption for the wakes of Mars and Venus, then $B_{rd}$ = $B_{\varphi d}$ = 0. Therefore, Equation 27 transforms in:

$$\bar{B} = \frac{\mu_0}{2R}\frac{j_0}{D}[1 - exp(-DR^2)]\widehat{\varphi_0} + B_{xd}\hat{\imath} \quad (28)$$

which is a helical field that winds around a cylindrical surface of radius R. Note that, regardless in which magnetic lobe the lines of the magnetic field are, they always turn in the clockwise direction -with respect to the planet-wake direction-, which is precisely the direction reported by Lundin et al. (2013). In the end, charged particles would be prone to follow the magnetic field lines and also turn in the clockwise direction.
Furthermore, following the Corollary II, let us consider the magnetic field given by Equation 28 with $B_{xd}$ constant. The Laplacian of B in cylindrical coordinates is:

$$\nabla^2 \bar{B} = \left[\frac{\mu_0}{2R^3}\frac{j_0}{D}(1 - exp(-DR^2)) - \mu_0 j_0 DR exp(-DR^2)\right]\widehat{\varphi_0} \quad (29)$$

And from corrollary II (equation 18) the vorticity is:

$$\bar{\omega} = \left[\frac{j_0}{2R^3 Dnq}(exp(-DR^2) - 1) + \frac{j_0}{nq}DRexp(-DR^2)\right]\widehat{\varphi_0} \quad (30)$$

which indicates that the vorticity declines very quickly with R, such that we would expect a very sharp maximum of vorticity close to the axis of the current. This can actually be observed in spacecraft data as we will see in Section 5.

2.3. Adaptation of the model to discrete data sets
In order to do a vortex analysis of a particular data set -as we do in Section 3-, we implement the ideas exposed in Sections 2.1 & 2.2, in particular, we apply Equation 18, which can be explicitly written -in Cartesian coordinates- as:

$$\bar{\omega} = -\frac{1}{\mu_0 nq}\left[\left(\frac{\partial^2 B_x}{\partial x^2} + \frac{\partial^2 B_x}{\partial y^2} + \frac{\partial^2 B_x}{\partial z^2}\right)i + \left(\frac{\partial^2 B_y}{\partial x^2} + \frac{\partial^2 B_y}{\partial y^2} + \frac{\partial^2 B_y}{\partial z^2}\right)j + \left(\frac{\partial^2 B_z}{\partial x^2} + \frac{\partial^2 B_z}{\partial y^2} + \frac{\partial^2 B_z}{\partial z^2}\right)k\right] \quad (31)$$

Given that we consider a steady state for Equation 18, for the current density we use Equation 12 taking $\partial E/\partial t = 0$

$$\bar{J} = \frac{1}{\mu_0}\left[\left(\frac{\partial B_z}{\partial y} - \frac{\partial B_y}{\partial z}\right)i - \left(\frac{\partial B_z}{\partial x} - \frac{\partial B_x}{\partial z}\right)j + \left(\frac{\partial B_y}{\partial x} - \frac{\partial B_x}{\partial y}\right)k\right] \quad (32)$$

Nevertheless, since the spacecraft measurements are non-continuous or discrete, we do the following approximations by finite differences:

$$\bar{\omega} \approx -\frac{1}{\mu_0 nq}\left[\left(\frac{1}{\Delta x^2} + \frac{1}{\Delta y^2} + \frac{1}{\Delta z^2}\right)\left(B_{x(m+1)} + B_{x(m-1)} - 2B_{xm}\right)i\right.$$
$$+ \left(\frac{1}{\Delta x^2} + \frac{1}{\Delta y^2} + \frac{1}{\Delta z^2}\right)\left(B_{y(m+1)} + B_{y(m-1)} - 2B_{ym}\right)j$$
$$\left.+ \left(\frac{1}{\Delta x^2} + \frac{1}{\Delta y^2} + \frac{1}{\Delta z^2}\right)\left(B_{z(m+1)} + B_{z(m-1)} - 2B_{zm}\right)k\right] \quad (33)$$

$$\bar{J} \approx \frac{1}{\mu_0}\left[\left(\frac{\Delta B_z}{\Delta y} - \frac{\Delta B_y}{\Delta z}\right)i - \left(\frac{\Delta B_z}{\Delta x} - \frac{\Delta B_x}{\Delta z}\right)j + \left(\frac{\Delta B_y}{\Delta x} - \frac{\Delta B_x}{\Delta y}\right)k\right] \quad (34)$$

We report the product nω as an approximation of the vorticity (also, from Equation 18). Given that the variations of n are small, the value of the product nω is mainly due to ω, so we consider that nω is a good proxy for ω. Explicitly, our approximation is the following:

$$\bar{\omega}n \approx -\frac{1}{\mu_0 q}\left[\left(\frac{1}{\Delta x^2} + \frac{1}{\Delta y^2} + \frac{1}{\Delta z^2}\right)\left(B_{x(m+1)} + B_{x(m-1)} - 2B_{xm}\right)i\right.$$
$$+ \left(\frac{1}{\Delta x^2} + \frac{1}{\Delta y^2} + \frac{1}{\Delta z^2}\right)\left(B_{y(m+1)} + B_{y(m-1)} - 2B_{ym}\right)j$$
$$\left.+ \left(\frac{1}{\Delta x^2} + \frac{1}{\Delta y^2} + \frac{1}{\Delta z^2}\right)\left(B_{z(m+1)} + B_{z(m-1)} - 2B_{zm}\right)k\right] \quad (35)$$

We highlight that the two latter equations are only dependent on the magnetic field values.

## 3. Comparison to Venus and Mars observations

In this Section, we present the results of the application of our model to magnetic field data of three single Venus wake crossings by the VEX spacecraft: 22 and 23 August 2006 and 14 December 2006; and a single Mars wake crossing by the MGS spacecraft: 15 December 2005. In all cases, the coordinate system used is centered at the planet. that is, the positive x-axis points sunwards, the positive y-axis points opposite to the orbital motion of the planet and the positive z-axis is perpendicular to the xy plane and points north of the ecliptic plane.

### 3.1. Vorticity

We obtained values of vorticity in the wake and downstream sheath with Equation 35 and identified several peaks of vorticity, which suggest several vortices. Vorticity profiles can be seen in the lower panels of Figure 1, where ω is plotted versus the planet- centric distance along the x direction. Wu et al. (2010) define a vortex as a vorticity tube surrounded by a non-rotational flow. The peaks of vorticity seen in Figure 1 fulfill this definition. Actually, the whole wake has non-zero vorticity, but regions with well-defined vorticity peaks could be understood as vortex cores.

### 3.2. Electric current density

We have also calculated the current density with the magnetic field data measured by VEX and MGS for the four wake crossings applying Equation 34. In the upper panel of Figure 1, the current density is plotted versus the x coordinate. In the data sets studied there are several current density peaks that, in general, coincide with the vorticity peaks as it can be seen in each pair of figures.

### 3.3. Vorticity from the wake model

In Figure 1, we have also calculated the theoretical vorticity for the three Venus cases (a, b & c) with the vortex model presented in Section 2.2, i. e., Equation 30. For that purpose, we have used position and current density data for each day. The resulting curves (peaks) appear in red in the lower panels. The theoretical curves assume that the vortices are symmetric with the wake axis. In comparison with the model, the vortex observed on day 23 August is almost axisymmetric, while the vortices on days 22 August and 14 December deviate from the axis of the wake exhibiting a high variability in its orientation with time.

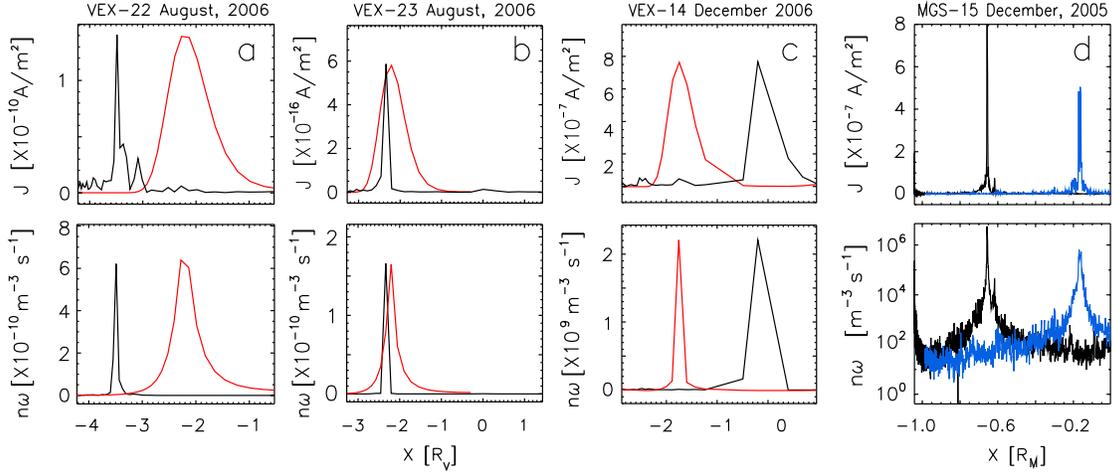

Figure 1: Examples of vortices identified in the wakes of Venus and Mars. The figures show the current density and the vorticity (times plasma density) along the wake from Equations 34 and 35. Panels (a) to (c) correspond to VEX data. The last set of panels (d) correspond to MGS data, where the black and blue trends correspond to inbound and outbound data with respect to the wake. For the Venus cases (a, b and c), we have overplotted the theoretical model curve (red curves) derived from Equation 21 (upper row) and Equation 30 (lower row) applied to each data sets. For comparison purposes, curves have been normalized to the maximum current density values. The best match between both curves is obtained for day 23, given that the vortex detected on this day is almost symmetric with respect to the wake axis.

4. Discussion

From our analysis, we conclude that vortices in non-collisional magnetized plasmas occur if there is a non-zero electric current (see Theorem I and its corollary), albeit the opposite is not true, since there could be an effective current ($J \neq 0$) and still $\nabla \times J = 0$, but no vorticity at all. In consequence, a necessary condition for the existence of a vorticity field is an electric current, which $\nabla \times J \neq 0$. The latter also allows us to conclude that the vortex detected by Lundin et al. (2013) in the wake and sheath of Venus requires a rotational electric current that generates it.

Furthermore, in non-collisional plasmas, electric currents can be generated by the difference in the speeds of the plasma species, i. e., $u_p \neq u_e$. Similar velocity gradients have been studied by Perez-de-Tejada and Durand-Manterola (1996) and Durand-Manterola (1997) who show that the mean velocity of charged particles in plasma velocity gradients is a function of the particle's mass. However, these gradients are not considered in this work because they occur close to the planet, while the Venus vortex we discuss in this work is further away in the wake. Nevertheless, some of the similarities between the Mars and Venus wakes suggest that both share similar generation mechanisms.

Our analysis shows a clear connection between the electric current and the vorticity -and thus, vortices- as seen either with VEX or MGS data in Figure 1 (a to d). In particular, we highlight that, to our knowledge, no other work -besides this current

work- has reported a phenomenon in Martian wake, similar to the one reported by Lundin et al. (2013).

The theorems we have introduced in Section 2, are useful vortex identification tools. In particular, through Theorem II and its corollary, we show that the vorticity can be expressed as a function of the magnetic field. In order to have an effective vorticity ($\omega \neq 0$), it is necessary that either $\Box^2 B \neq 0$ or $\nabla^2 B \neq 0$, which suggests that the mechanism that generates vortices in non-collisional plasmas is the torsion of the magnetic field, generated by the current, that forces the plasma to rotate producing a vortex.

It is worth noting that our approach is somewhat different from that of Banks et al. (1981), who use the concept of generalized vorticity, a composite of the vorticity $\nabla \times \vec{u}$ (we have used along the current work) and the gyrofrequency, $q\vec{B}/m$. According to Banks et al. (1981), the balance between these two terms define four different regimes in a collision less plasma described in terms of the standard momentum equation valid for each of its charged components. To a first approximation and in an effort to contextualize our analysis within this frame, our work, or at least the cases (Figure 1) we consider, fall in the Ordinary Fluid Regime given that $\nabla \times \vec{u} \gg q\vec{B}/m$. Note that the gyrofrequency of electrons and protons in this environment is $\sim 10^3$ s$^{-1}$ and $\sim 1$ s$^{-1}$, respectively; while the vorticity is, in all cases, $> 10^3$ s$^{-1}$. Banks et al. (1981) state that, in this case, Kelvin's circulation theorem holds true for each species and that electrodynamic effects may still be present since it is possible for currents to create magnetic vector potentials and do not violate the assumption. This latter statement is clearly demonstrated in the current work.

One more detail is that, in the cases that we have analyzed in this work, the axis of the vortex does not always coincide with the axis of the wake, as can be seen in Figure 1, where the main vorticity peak is found in the sheath and not in the wake. Also, one can see more than one vortex, which is the case of Mars, where we identified two in the wake (Figure 1 d). The presence of two or more vortices could be explained in two ways: one possibility is that the spacecraft has passed through a region where the structure of the vortex is bent and crosses it at different locations. Actually, another possibility could be that there are two or more independent vortices due to the presence of a magnetic field with a complex structure where there are two or more electric currents. The comparison between the VEX data sets of the three wake crossings studied shows that the Venus wake vortex is very dynamic, changing from one day to the next, which also indicates that the vortex is very sensitive to the surrounding environment conditions. Anyway, part of these variations may be due to the slight change of direction that the VEX orbit has from one day to the next. Apparently, this statement contradicts the steady state assumption used in the derivation of our theorems, but it must be considered that every wake crossing is fast enough ($\sim$ few minutes) such that our assumptions are held. Of course, on a longer time scale ($\sim$days), the steady state assumption is no longer valid, and we would need to apply only Theorem II and not its corollary.

Also derived from the analysis of VEX magnetic field data (from 2006 to 2014), Chai et al. (2016) reports a global induced magnetic field -different to the draped field- whose distribution corresponds to a cylindrical shell around the magnetotail. This extra field turns in the same direction reported by Lundin et al. (2013) (see Figure 4 of Chai et al. (2016)). They also report that this field seems to be located from 0.5 RV to -at least-2.5 RV. This cylindrical field, designated by Chai et al. (2016)), looping field, supports our statement, since, in terms of our model, it corresponds to the field generated by the electric current around the axis of the magnetotail.

In the cases of Venus and Mars where the wake plasma is overall superalfvenic, one could disagree with some of the ideas exposed in this work arguing that, in such environment, it is the plasma which drags the field and not the other way around. Nevertheless, this objection can be disregarded if we consider that the force exerted by the helical field is perpendicular to the direction of the flow where there is no effective plasma kinetic pressure. Actually, the only pressure that opposes the magnetic pressure is the thermal pressure, which is indeed smaller than the magnetic pressure itself (Dubinin et al., 2013).

Finally, we wish to highlight that to our knowledge -so far-, besides the current work, there is no other work or model in the literature that explains the driving mechanism behind the vortices observed in the Venus' wake. Furthermore, the detection of vortical structures in their wakes, in Venus and Mars, is important, because it suggests that such structures may also be present in the plasma wake of other planetary bodies, where there is a relevant interaction between the solar wind and the ionospheric plasmas, for example in the case of cometary bodies and the dwarf planet Pluto.

5. Conclusions

The conclusions of this work are the following:

· In non-collisional magnetized plasmas, the presence of a vortex implies the existence of an electric current tube.

· The mechanism that produces this type of vortices is the mutual interaction between the electric current and the magnetic field. The current induces a magnetic field that twists the original field. This latter field twists the plasma in a helical trajectory: the vortex.

· In the observed data, the peak of electric current coincides, in position, with the peak of vorticity, supporting our first conclusion.

· Vortices exhibit important changes from one day to the other, from which we conclude that they are quite dynamic structures and therefore very sensitive to the surrounding plasma changes due to the solar activity.

· Mars has wake vortices similar to the vortices in Venus.

· Several vortices can coexist at the same time at different locations of the wake and the sheath.

Acknowledgements
AF and HPT were supported by DGAPA/UNAM through the projects IN105818 and IN103119, respectively. All the data used in this work were downloaded from CLWEB http://clweb.irap.omp.eu/cl/clweb.php, and AMDA http://amda.irap.omp.eu/.